\documentclass[amsmath,amssymb,aps,twocolumn,nofootinbib,preprintnumbers]{revtex4}
\usepackage[dvips]{graphics,color}
\usepackage{latexsym}
\usepackage{graphicx}
\newcommand{\ssection}[1]{\noindent{\it #1}\,:}
\begin{document}

\preprint{KUNS-2092}

\title{No de Sitter invariant vacuum in massive gravity theory with ghost}

\author{
Keisuke Izumi\footnote{e-mail:
ksuke@tap.scphys.kyoto-u.ac.jp}
and 
Takahiro Tanaka\footnote{e-mail:
tama@scphys.kyoto-u.ac.jp}\\~}

~\\

\address{Department of Physics, Kyoto University, Kyoto 606-8502, Japan}

\begin{abstract}
In this letter we point out that the massive gravity theory with a graviton ghost mode  
in de Sitter background cannot possess a de Sitter invariant vacuum state. 
In order to avoid a negative norm state, 
we must associate the creation operator of the ghost mode 
with a negative-energy mode function instead of a 
positive-energy one as the mode function. 
Namely, we have to adopt a different procedure of quantization for a
 ghost. 
When a theory has a symmetry mixing a ghost mode with ordinary non-ghost modes, 
the choice of a ghost mode is not unique. 
However, quantization of a ghost is impossible without
specifying a choice of ghost mode, which breaks the symmetry. 
For this reason, the vacuum state cannot respect the symmetry.
In the massive gravity theory with a graviton ghost mode
in de Sitter background, the ghost is the helicity-0 mode 
of the graviton. 
This ghost mode is mixed with the other helicity graviton modes
under the action of de Sitter symmetry.
Therefore, there is no de Sitter invariant vacuum in such models.
This leads to an interesting possibility that 
non-covariant cutoff of the low energy effective theory may naturally arise.
As a result, 
the instability due to the pair production of a ghost and normal non-ghost
 particles gets much milder and 
that the model may escape from being rejected.
\end{abstract}

\maketitle
\ssection{Introduction}

Discovery of the accelerated expansion of the universe is one of the
hottest current topics in cosmology. 
To explain this observation, various ideas such as 
quintessence, phantom-field, $f(R)$-gravity, scalar-tensor theory and so
on, are proposed~\cite{darkenergy}.  
All of them can be classified as 
modifications in the spin-0 sector.
Another interesting alternative to pursue will be  
the modification in the spin-2 sector, which has been studied less
extensively so far~\cite{cedric1,cedric2,koyamareview}. 

The simplest model of gravity theory with modified spin-2 sector is 
the massive gravity theory~\cite{FP}, 
which has the terms quadratic in metric perturbation.
When we consider small perturbations around a given background
in a model of modified gravity theory, 
in many cases the model is reduced to the massive gravity theory 
with higher-order coupling terms.   
In order to explain the current accelerated expansion of the universe 
by considering this type of models, 
naively the value of the graviton mass $m$ should be comparable 
to the present value of the Hubble constant $H$.
However, when the mass of the graviton is in the range $0<m^2 <2H^2$, 
the helicity-0 mode of the graviton becomes a ghost mode
in de Sitter background~\cite{Higuchi}.
Here a ghost mode means a mode with the overall signature of  
the action being opposite. 
Though there is no proof of absence, 
no ghost-free model in which the effect of spin-2 sector 
makes the cosmic expansion accelerated has been 
found~\cite{effective1,effective2,Koyama1,Koyama2,Charmousis,Charmousis2,Carena,Izumi}.

If a ghost mode is quantized in the usual manner,
a negative norm state emerges.
In order to avoid the negative norm state, 
One possible prescription is to exchange the mode functions attached to the creation 
and annihilation operators~\cite{Cline}. 
As a result, the excitations of the ghost particles have negative
energy. 
Excitations with negative energy is not unusual in cosmology, e.g. tachyon. 
A distinguishing feature of the ghost, however, is in that 
a ghost mode has negative energy even for the high momentum states, 
which is thought to lead to a disaster. 
In the flat spacetime, 
if a ghost mode couples to normal particles, 
whose excitation energy is positive, 
pair-creation from the vacuum is possible 
because the total energy of the ghost and the normal particles 
can be zero. 
The probability of the reaction which is obtained by boosting 
the original one by a Lorentz transformation
is the same.
The total number density and the total energy density of the created normal particles 
are given by integrals over the whole 3-dimensional momentum space, and 
therefore they instantaneously become infinite
due to the divergent UV-contribution, 
as long as the pair-creation process for given momenta has 
a finite probability no matter how small it is.  
When the model has the Lorentz symmetry, 
it is natural that the UV-cutoff scale (if it exists) is to be imposed 
in the Lorentz invariant manner in the 4-dimensional momentum space.
Then, the 3-dimensional integrals mentioned above 
have no chance to have a natural cutoff, 
and the UV-divergence cannot be avoided.
This means that such a model should be rejected. 

However, the same story does not apply to 
the massive spin-2 field with a ghost mode in de Sitter background. 
We will show that, in the massive gravity theory with a ghost mode,
a de Sitter invariant vacuum does not exist. 
Thus, any vacuum state necessarily breaks the de Sitter symmetry.  
This leads to an interesting possibility that 
a cutoff in 3-dimensional momentum space may naturally arise as a consequence of 
the symmetry breaking. 
If the model is cut off at a finite magnitude of the 3-dimensional momentum, 
the total number density and the total energy density of 
created normal particles remain finite, 
so that the model may escape from being rejected. \\

\ssection{Symmetry breaking due to ghost}

Usually the creation operator of a mode is associated 
with a positive-energy mode function. 
However, if we apply this prescription to 
a model with a ghost, negative norm states appear.
One possible way to avoid this pathological situation 
will be to assign a negative-energy mode function to the creation operator.
However, when the model has a symmetry which mixes the ghost mode 
with ordinary non-ghost modes,
there are infinitely many variations in identifying the ghost mode, 
for which we flip the roles of positive and negative-energy mode functions. 
Each choice gives a different vacuum state, and hence each vacuum state 
does not respect the symmetry. 

Before we discuss the the massive gravity theory 
in de Sitter background, 
we show a simple example to demonstrate how 
the choice of the ghost mode breaks the symmetry.
We consider a model which consists of two degrees of freedom. 
The action of the model is given by 
\begin{eqnarray*}
S = {1\over 2}\int dt \left[
(\partial_t  \phi)^2 - (\partial_t \psi)^2-\omega^2(\phi^2-\psi^2)
\right]. 
\end{eqnarray*}
A ghost mode is represented by $\psi$, 
while the ordinary non-ghost mode by $\phi$. 
This action is invariant under the transformation 
\begin{eqnarray*}
\phi \to \phi_\theta = \phi \cosh\theta + \psi \sinh\theta, \\
\psi \to \psi_\theta = \psi \cosh\theta + \phi \sinh\theta, 
\end{eqnarray*}
where $\theta$ is a parameter of the transformation. 
Namely, the action can be also written as
\begin{eqnarray*}
S = {1\over 2}\int dt \left[
(\partial_t  \phi_\theta)^2 - (\partial_t \psi_\theta)^2
 -\omega^2(\phi_\theta^2-\psi_\theta^2)
 \right].
\end{eqnarray*}
This transformation obviously mingles ghost and non-ghost degrees of 
freedom. 

If we assign the creation operator of the ghost mode $\psi$
to the positive-energy mode function proportional to
$\exp(-i\omega t)$, states with a ghost particle become 
negative norm states. To avoid this pathology, as we stated earlier, 
we associate the negative-energy mode function with the creation operator.
However, the action $S$ itself does not uniquely specify the choice of the ghost mode. 
$\psi_\theta$ with any value of $\theta$ can be equally good as a ghost mode. 
If the original variables $\psi$ 
is selected as a ghost degree of freedom, $\phi$ and $\psi$ are expanded as 
\begin{eqnarray}
\phi= {1\over \sqrt{2\omega}}
  \int d\omega \left(a_{\phi} \exp(-i\omega t) +a_{\phi}^\dagger
		\exp(i\omega t)
     \right), \nonumber\\
\psi= {1\over \sqrt{2\omega}}
  \int d\omega \left(a_{\psi} \exp(i\omega t) +a_{\psi}^\dagger
  \exp(-i\omega t)\right),
\label{expand} 
\end{eqnarray}
naturally introducing the associated creation and annihilation
operators. 
The same is true for $\psi_\theta$ with 
any value of $\theta$. We can expand $\psi_\theta$ and $\phi_\theta$ as  
\begin{eqnarray*}
\phi_\theta=  {1\over \sqrt{2\omega}} 
   \int d\omega \left(a_{\phi,\theta} \exp(-i\omega t)
		 +a_{\phi,\theta}^\dagger 
      \exp(i\omega t)\right), \nonumber\\
\psi_\theta=  {1\over \sqrt{2\omega}}
   \int d\omega \left(a_{\psi,\theta} \exp(i\omega t) +a_{\psi,\theta}^\dagger
   \exp(-i\omega t)\right),
\end{eqnarray*}
instead of (\ref{expand}). 
Then, the relations between the creation and annihilation operators are
\begin{eqnarray}
a_{\phi,\theta} = a_{\phi} \cosh\theta + a_{\psi}^\dagger \sinh\theta,\nonumber \\
a_{\psi,\theta} = a_{\psi} \cosh\theta + a_{\phi}^\dagger \sinh\theta.
\label{Bog}
\end{eqnarray}
We can see that there is the mixing between the creation and annihilation operators.
The vacuum state $ \left| 0\right\rangle$ 
associated with the original variables $\phi$ and $\psi$
is defined by the conditions 
$a_{\phi} \left| 0\right\rangle =0$ 
and $a_{\psi} \left| 0\right\rangle =0$.  
The vacuum state $ \left| 0\right\rangle_\theta$ is also defined in the 
same manner. Then, we have 
\begin{eqnarray*}
a_{\phi,\theta} \left| 0\right\rangle 
=a_{\psi}^\dagger \sinh\theta \left| 0\right\rangle
\neq 0 .
\end{eqnarray*}
Namely, the vacuum $\left| 0\right\rangle$ 
is not identical to the vacuum $\left| 0\right\rangle_\theta$.  
This means that the symmetry is broken in each vacuum state\footnote{
We can promote $\phi$ and $\psi$ to scalar fields.   
In the same way the symmetry parameterized by $\theta$
is spontaneously broken. 
Then, Nambu-Goldstone theorem applies and it indicates the 
existence of a massless excitation. 
Although this model does not have any massless one-particle state, 
it is not a contradiction. 
Such a massless state can be easily constructed by considering 
multi-particle states involving both ghost and non-ghost particles.
}, 
and there is no vacuum state which respects the symmetry\footnote{
Of course, the state which is obtained by 
$\int_{-\infty}^\infty d\theta\, \vert 0\rangle_\theta$ is 
manifestly invariant under the current symmetry operation.  
However, this state is obviously unnormalizable, and 
also it is a superposition of highly excited states 
from any vacuum state $\vert 0\rangle_\theta$. 
}.

In contrast, we can consider a model similar but 
without a ghost, whose
action is given by 
\begin{eqnarray*}
S ={1\over 2} \int dt \left[
(\partial_t  \phi)^2 + (\partial_t \psi)^2
 -\omega^2(\phi^2+\psi^2)
 \right].
\end{eqnarray*}
In this case the counterparts of the relations (\ref{Bog}) do not
contain the creation operators. Hence, all vacua parameterized by
$\theta$ are identical. In this case therefore there exists a vacuum state 
which respects the symmetry. \\

\ssection{Spontaneous symmetry breaking of massive graviton}

Now, we turn to the discussion about the massive spin-2 field with a
helicity-0 ghost mode living in de Sitter background.
The background metric is given by  
\begin{eqnarray*}
ds_{(0)}^2=\frac{1}{(H \eta)^2}(-d\eta^2 + d{\bf x}^2), 
\end{eqnarray*}
where $H$ is the expansion rate of the universe. 
The action of the massive gravity theory is written as~\cite{FP}
\begin{eqnarray}
&&S = {M_{pl}^2\over 2}\int d^4 x \sqrt{-g} \cr
&&\qquad\qquad\times \left(R -6H^2 - \frac{m^2}{4}(h^{\mu\nu}h_{\mu\nu}
		      -h^2)
\right), \label{massive}
\end{eqnarray}
where $h_{\mu\nu}\equiv g_{\mu\nu}-g^{(0)}_{\mu\nu}$ and
$g^{(0)}_{\mu\nu}$ is 
the background metric. We denote the Planck mass, the Ricci curvature 
and the graviton mass by $M_{pl}$, $R$ and $m$, respectively. 
This action has the de Sitter symmetry. 
When there is no matter source, 
the graviton satisfies transverse-traceless conditions, and 
it has five polarizations. One of them is the helicity-0 mode. 
Helicity-1 and helicity-2 modes have two polarizations, respectively. 
In the terminology 
frequently used in the context of cosmological perturbation,  
this helicity-0, 1, and 2 modes are called 
3-dimensional scalar, vector and tensor modes.  
At the linear level different helicity modes decouple from 
each other. 
However, the action of the de Sitter group which leads to the variation 
of the time slicings mingles 
different helicity modes\footnote{
In the Minkowski background, the helicity-0 base of the spin-2
(=transverse traceless) tensor is expressed as
$e_{\mu\nu}^{(0)}
=2e^{(0)}_{\mu}e^{(0)}_{\nu}-e^{(1)}_{\mu}e^{(1)}_{\nu}
-e^{(2)}_{\mu}e^{(2)}_{\nu}$, 
where $e^{(0)}_{\mu}=m^{-1} (k_1,k_0,0,0)$, $e^{(1)}_{\mu}=(0,0,1,0)$,
$e^{(2)}_{\mu}=(0,0,0,1)$ and $k_\mu=(k_0,k_1,0,0)$ is 4-momentum.
If we take another frame connected by a Lorentz transformation,
$e^{(i)}_{\mu}$s changes, and hence $e_{\mu\nu}^{(0)}$ 
changes as well. As a result, 
the different helicity bases of the spin-2 tensor also enters 
into $e_{\mu\nu}^{(0)}$.
Since the Minkowski case 
is just a special limit, $H\to\infty$,  
of the de Sitter case,
the mixture of different helicity bases occurs also in de Sitter background 
by the action of the de Sitter group.
}.

The metric perturbation corresponding to the 
helicity-0 mode is explicitly written as
\begin{eqnarray*}
 h_{\mu\nu}=\left(
  \begin{array}{cc}
    s& -{\nabla_i\over k^2}\left(\partial_\eta
    -{2\over \eta}\right)s\cr
    * & \cdots 
    \end{array}
\right), 
\end{eqnarray*}
using a master variable $s$. 
Here we omitted the lengthy space-space component.  
The quadratic effective action for this mode is given by~\cite{Higuchi}
\begin{eqnarray}
&& S_{ghost}=\int dk^3\int 
  {M_4^2 m^2(m^2-2H^2)\over k^4}\cr 
 &&\qquad\qquad\times s_k
  \left[\Box_{flat}+{2\over \eta}\partial_\eta 
     -{m^2\over H^2}\right]s_k\, d\eta,  
\label{effective}
\end{eqnarray}
where $s_k(\eta)$ is the spatial Fourier component of $s$. 
For $m^2<2H^2$, the signature of this action becomes 
negative and the mode becomes a ghost. 
Since the original action (\ref{massive}) has the de Sitter symmetry,
there are various equivalent ways of choosing the time-constant 
surfaces. Depending on how we foliate the time-constant 
surfaces, the definition of helicity-0 mode varies, 
which leads to ambiguity in specifying the ghost
degree of freedom.
Namely, 
the meaning of the helicity-0 mode depends on the choice of the 
spacetime foliation, since the different helicity modes are 
mixed under the action of the de Sitter group. 
As we have seen above, in order to avoid the negative norm state, 
the ghost degree of freedom must be specified.
This means that we need to fix spacetime foliation 
in quantizing a massive graviton with $0<m^2<2H^2$. 
A different choice of the ghost or equivalently a different 
choice of the foliation leads to a different vacuum. 
Therefore the vacuum state necessarily breaks de Sitter symmetry.

When we consider small fluctuations around a fixed background, 
most of the models of modified gravity theory
reduce to the massive gravity theory with non-linear coupling terms.
Hence, how to perturbatively quantize such models is the same 
as the massive gravity theory. 
Then, we need to specify a preferred frame to fix the vacuum state, 
and the vacuum state breaks de Sitter invariance. \\

\ssection{Discussion}

A ghost can be made harmless once we introduce a non-covariant cutoff in the 
3-dimensional momentum space. 
In the above discussion, 
we have shown that the models of massive gravity in de 
Sitter background do not possess a de Sitter invariant vacuum state 
when there is a helicity-0 ghost. 
Is there any implication of this fact to the issue of non-covariant
cutoff?  
From the point of view of the low energy effective theory, 
a natural cutoff will arise from the validity range of the model. 
In the case of massive gravity, it is well known that the models 
become strongly coupled at a relatively long distance scale~\cite{strong1,strong6,strong2,strong3,strong4,strong5}. 
Here, one crucial point is whether the cutoff that appears in such a way 
is covariant or not. 
If the cutoff is non-covariant, 
the processes of pair production with largely boosted momenta 
might be in the strongly coupled regime, and hence they are 
outside the validity range of perturbative expansion. 
Then, we do not know how to compute the probability of such processes. 
Thus, we cannot conclude that the total pair production rate is 
infinitely large, in contrast to the case of a ghost in the 
flat background. The total pair production rate might be moderately small. 
We will not pursue a rigorous proof of it here, but we think it quite 
unlikely that the cutoff is covariant under the situation that a vacuum state chooses 
a preferred frame. 
In fact, the effective action for the helicity-0 mode shown in
Eq.~(\ref{effective}) 
manifestly depends on the magnitude 
of the 3-momentum $k$. 
Since we took different quantization procedure for $s$ (helicity-0), 
we have to treat the propagator of $s$ differently from the other
modes (helicity-1,2) in the Feynman rule. 
As a result, Feynman diagram does not respect the de Sitter invariance. 
As we 
determine the strong coupling condition by comparing the tree-level amplitude 
with the loop corrections, this condition 
cannot be covariant, either. 

From the expression~(\ref{effective}), we can read that the expectation value 
of the amplitude of fluctuations $\langle s_k^2\rangle$ 
becomes very large for large $k$.  
This means that models of massive gravity have a tendency 
to be strongly coupled even at a relatively long wavelength. 
For the known examples of massive gravity in the flat background, 
the strong coupling scale is given by $\sim (M_{pl} m^{n-1})^{1/n}$
with $n=3$ or 5~\cite{strong2}. 
Thus, it seems natural to assume that 
the cutoff energy scale is significantly lower than the Planck scale. 
To determine the condition for the strong coupling, 
the higher order interaction terms, 
which are not contained in the action (\ref{effective}), are necessary. 
Those terms are model-dependent. 
We defer the evaluation of strong coupling scale 
in a specific model like DGP model~\cite{DGP} 
in de Sitter background to the future work~\cite{koyama3}. 
To conclude, our message is that the modified gravity theory with a graviton 
ghost may avoid the disastrous instability thanks to the 
spontaneous breaking of the de Sitter symmetry. 
The three-dimensional cutoff may naturally arise as 
the strong coupling scale. 
Therefore the existence of a ghost may not directly indicate the
inconsistency of the model.

In our previous paper~\cite{Izumi}, 
which discusses the DGP braneworld model having two branes,  
we showed that, if the model parameters, such as the tensions of the
branes, are continuously changed, 
the ghost degree of freedom can be transfered from the spin-2 sector
to the spin-0 sector. Roughly speaking, the spin-0 ghost 
corresponds to the brane separation. 
After the ghost degree of freedom shifts to the spin-0 mode,
the ambiguity of the spacetime foliation does not affect the 
choice of the vacuum any more. 
Namely, the ghost degree of freedom is uniquely specified. 
Then it seems that one can choose a vacuum state which respects 
the de Sitter symmetry.
However, in this model the mass of this spin-0 ghost mode is $m_0^2 =-4H^2$
(or smaller when a bulk scalar field is introduced). 
If $m_0^2 >-4H^2$, the ghost is on the spin-2 graviton side. 
 As is known well, 
de Sitter invariant vacuum of a free scalar field
does not exist for $m_0^2 \le 0$~\cite{Allen,Vachaspati,Kirsten},
and hence the vacuum necessarily breaks the de Sitter symmetry again.

The equivalence theorem states that 
the helicity-0 action of the massive gravity theory~(\ref{massive})
can be rewritten as a spin-0 mode in the high energy limit. 
Based on it, 
one may claim that 
the de Sitter symmetry must be preserved in the high energy limit. 
If this argument is correct, our statement 
presented above should have some errors.   
However, the equivalence theorem does not apply to 
the present case for the following reason. 
Here, the term "high energy limit" means that
the momentum of one external line of the 
spin-2 helicity-0 mode is large in the center of mass frame of 
another external line.
Since we are concerned with the processes which are obtained by 
boosting a low energy process,  
this momentum is, of course, small.  
Therefore the equivalence theorem does not apply to the present case and 
our argument that the cutoff condition due to the strong coupling 
will be non-covariant does not conflict with the theorem.

\acknowledgements
The authors thank Takashi Nakamura for his valuable comments and continuous
encouragement.
TT is supported by Grant-in-Aid for
Scientific
Research, Nos. 16740141 
and by Monbukagakusho Grant-in-Aid
for Scientific Research(B) No.~17340075. 
This work is also supported in part by the 21st Century COE ``Center for
Diversity
and Universality in Physics'' at Kyoto university, from the Ministry of
Education,
Culture, Sports, Science and Technology of Japan
and also by the Japan-U.K. Research Cooperative Program
both from Japan Society for Promotion of Science.

\end{document}